\documentclass[12pt]{article}
\usepackage{epsfig,psfrag,cite,float,amssymb}
\def \beq{\begin{equation}}
\def \eeq{\end{equation}}
\def \bea{\begin{eqnarray}}
\def \eea{\end{eqnarray}}
\def \ben{\begin{enumerate}}
\def \een{\end{enumerate}}
\def \bit{\begin{itemize}}
\def \eit{\end{itemize}}

\def \branch{{\cal B}}
\def \gev{{\hbox{GeV}}}

\def \cl#1{{#1\%\ \mathrm{C.L.}}}

\def \eq#1{Eq.~(\ref{#1})}

\def \fig#1{Fig.~\ref{#1}}

\def \rf{Ref.~\cite}
\def \rfs{Refs.~\cite}
\def \sec{Sec.~\ref}

\def \a{\alpha}
\def \b{\beta}
\def \D{\Delta}
\def \g{\gamma}

\def \d{\delta}

\def \m{\mu}
\def \n{\nu}

\def \p{\pi}

\def \s{\sigma}

\def \oi{{\cal O}}

\begin{document}
\thispagestyle{empty}
\setcounter{page}{0}
\begin{flushright}
ZU-TH 8/01\\
LMU-xx-/01\\
DESY 01-128\\
\end{flushright}
\vspace*{\fill}
\begin{center}
{\Large \bf
Complex flavour couplings in supersymmetry and unexpected 
$\mathbf CP$-violation in the decay $B \to \phi K$}

\vspace{2em}

\large
E. Lunghi\footnote{E-mail: lunghi@mail.desy.de} and
D. Wyler\footnote{permanent address:
Theoretische Physik, Universit\"at Z\"urich
Winterthurerstrasse 190, CH-8057, Z\"urich. 
E-mail: wyler@physik.unizh.ch}\\
\normalsize
Deutsches Elektronen Synchrotron, DESY, \\
Notkestrasse 85, D-22607, Hamburg, Germany \\ 

\large 

\vspace{0.5cm} 

\end{center}
\vspace*{\fill}

\begin{abstract}
Complex flavour couplings (off-diagonal mass terms) in the squark
sector of supersymmetric theories may drastically alter both the rate and
the $CP$-violating asymmetry of certain $B$-meson decays. We consider the
effects of couplings that induce $b \to s$ transitions and lead
to final state with strangeness one. We investigate the bounds that
must be satisfied by the new terms and explore the possible
implications on direct and mixing induced asymmetries in the charged
and neutral $B \to J/\Psi K$ and $B \to \phi K$ decays.
\end{abstract}

\vspace*{\fill}


\newpage

\section{Introduction}
\label{intro}
Certain decays of $B$-mesons are expected to shed light on the
mechanism of $CP$ violation and, more generally, on new physics.
Indeed, the first surprising results on the $CP$-asymmetry $a_{\psi
K}$ in the decay $B\to J/\Psi K$ raised the hope for a first new
physics signal. Meanwhile, the value of the measured asymmetry has
changed considerably and the present world average~\cite{beta}
\beq 
a_{\psi K}^{WA} = 0.79 \pm 0.12 
\eeq
almost coincides with the standard model expectation~\cite{smod}
\beq
a_{\psi K}^{SM} \simeq 0.70 \pm 0.10 \; .
\eeq
Although this came as a disappointment, it reinforced the view that
one should be prepared, both on the theoretical and the experimental
side, for unexpected observations. In fact, the small error of the
above world average and the improvements that are expected during the
next years will allow to detect even small deviations from the
standard model predictions.

There are of course many new physics scenarios. On the one hand, it is
possible to describe their signatures in very general terms (for
recent expositions, see e.g. \rfs{fun,nir,manflei1}). On the other
hand, one can propose a specific new physics model and investigate its
consequences. The latter approach was the topic of a huge body of
work. For what concerns supersymmetric models, emphasis was given
recently to the so-called minimal flavour violating models
(MFV)~\cite{mfv} and their possible variants~\cite{variants}.

In virtually all new physics models, new non-standard fields are
introduced and, therefore, new complex (i.e. with $CP$-violating
phases) couplings appear. This is well known in supersymmetry to which
we turn for definiteness. In supersymmetric models there are several
classes of phases. Those in the $\m$ and flavour diagonal $A$ terms do
not contribute to flavour changing processes but are strongly bound by
the electric dipole moment of the neutron and other
particles~\cite{b}. The phases of the Yukawa couplings are the same as
in the standard model. If these are the only phases and flavour
changing couplings of a model, the latter belongs to the class of MFV
supersymmetric models that exhibit many analogies to the standard
model~\cite{ mfv}. Therefore, the most interesting phases are those in
the squark and slepton mass matrices. Their flavour diagonal elements
are either real by definition (in the $LL$ and $RR$ sectors) or small
(in the $LR$ and $RL$ ones) as pointed out before (restrictions from
electric dipole moments).  Therefore the flavour changing elements
(i.e.  $(m^2)_{23}^{u,d}$ and $(m^2)_{13}^{u,d}$, in the sectors $LL$,
$RL$, etc.) are the most interesting ones and we will focus on their
effects. They contribute mainly through loop diagrams with internal
gluinos and charginos.  Since we are interested in contributions to
Wilson coefficients that are already quite large in the
SM~\footnote{Contributions to the small electroweak penguin
coefficients are usually negligible because they do not appreciably
change the standard model predictions.}(i.e. the coefficients of the
QCD penguin and chromo-magnetic dipole moment operators), we will only
discuss gluino loops; in fact, they tend to dominate over the
corresponding chargino ones.

Clearly this is not the first analysis of such SUSY diagrams. Their
impact on flavour physics is known since a long time~\cite{dnw}, and there are
numerous recent investigations. To clarify the new aspects provided in
this paper, we review briefly some of the most recent efforts.

In \rf{gluino} a first comprehensive analysis of the gluino exchanges
was given. The systematics of the perturbative expansion was
investigated in \rf{bghw}, which also included an analysis of the mass
insertion approximation: the latter was found to be sufficient in many
cases and we will use it again in this study.
In \rf{bhg} the bounds from the rare decay $b \to s \gamma$ were
derived; however, the various couplings were assumed to be real.
In \rf{troja}, the influence on the electromagnetic penguins (isospin
violating terms) and the consequences on $B \to K \pi$ decays were
studied; particular focus was given to the determination of the $CKM$
angle $\gamma$.  In \rf{fun} effects of new terms on direct
$CP$-violating asymmetries in charged decays were given, but the
possible values of the new coefficients were not investigated. In
\rf{manflei1} a completely general parameterization of new physics
effects in the decays $B \to \phi K$ and $B \to J/\Psi K$ was given;
however, no particular model was explicitely studied.
In \rf{bernabeu}, the case of left-right symmetric models were
investigated and it was found that, in these models, sizeable
deviations from the relation $a_{\psi K} = a_{\phi K}$ are possible.

In this paper, we consider the {\bf complex} $(2,3)$ entries in the
down squark mass matrix and work out their consequences for the time
dependent $CP$ asymmetries in the decays $B \to \phi K$ and $B \to
J/\Psi K$. In the standard model, these asymmetries are equal and are
a measure of the phase of the $B_d$-$\bar B_d$ mixing amplitude (which
in the SM is $2 \beta$). The complex couplings that we consider leave
the $B_d$-$\bar B_d$ mixing phase unchanged but have a significant
impact on the amplitudes of the decays. The $B \to J/\Psi K$ amplitude
is dominated by SM tree level contributions and, consequently, the
$CP$-violating asymmetry is hardly modified; on the other hand, there
is room for such an effect in the transition $B \to \phi K$. This
decay is easily accessible at the $B$-factory experiments~\cite{kafi}
and an investigation of the related $CP$ asymmetries is clearly
worthwhile. In the analysis we will concentrate on this decay.

This note is organized as follows. In \sec{sec2} we review the
effective Hamiltonian for $b\to s \bar s s$ transitions in generic
extensions of the standard model and give the explicit expressions of
the various $CP$ asymmetries. We also discuss the observables $\branch
(B\to X_s \g)$ and $\D M_{B_s}$ which are related to the $b\to s$
transition and provide interesting additional pieces of
information. In \sec{sec3} we present the SUSY model that we consider
and the explicit contributions to the Wilson coefficients. The
numerical analysis is presented in \sec{sec4}. The impact of the
complex SUSY parameters on the branching ratio and $CP$ asymmetry of
$B\to X_s \g$ is explored and we analyze the correlation between the
mixing-induced (and direct) $CP$ asymmetries in $B \to \phi K$ and the
$B_s$-${\bar B}_s$ mass difference. A brief summary of our results and
some comments are found in \sec{concl}.

\section{Effective Hamiltonian}
\label{sec2}
The effective Hamiltonian for $b\to s$ transitions can be written as
\beq
{\cal{H}}_{eff} = {G_F\over \sqrt{2}} V_{tb}^{} V_{ts}^* \left[ \sum_{i=1}^{10}
                  C_i (\mu) \oi_i  + C_{7\g} \oi_{7\g} + C_{8 g} \oi_{8 g} \right]
\label{ham}
\eeq
where ($e_{u,c}=2/3$ and $e_{d,s,b}=-1/3$) 
\bea
\oi_1 &=& (\bar c_\a b_\b)_{V-A} (\bar s_\b c_\a)_{V-A} \\
\oi_2 &=& (\bar c b)_{V-A} (\bar s c)_{V-A} \\
\oi_3 &=& (\bar s b)_{V-A} \sum_{q=u,d,s,c,b} (\bar q q)_{V-A} \\
\oi_4 &=& (\bar s_\a b_\b)_{V-A} \sum_{q=u,d,s,c,b} (\bar q_\b q_\a)_{V-A} \\
\oi_5 &=& (\bar s b)_{V-A} \sum_{q=u,d,s,c,b} (\bar q q)_{V+A} \\
\oi_6 &=& (\bar s_\a  b_\b)_{V-A} \sum_{q=u,d,s,c,b} (\bar q_\b q_\a )_{V+A} \\
\oi_7 &=& {3\over 2} (\bar s b)_{V-A} \sum_{q=u,d,s,c,b} 
                     e_q (\bar q q)_{V-A} \\
\oi_8 &=& {3\over 2} (\bar s_\a b_\b)_{V-A} \sum_{q=u,d,s,c,b} 
                     e_q (\bar q_\b q_\a)_{V-A} \\
\oi_9 &=& {3\over 2} (\bar s b)_{V-A} \sum_{q=u,d,s,c,b} 
                     e_q (\bar q q)_{V+A} \\
\oi_{10} &=& {3\over 2} (\bar s_\a  b_\b)_{V-A} \sum_{q=u,d,s,c,b} 
                     e_q (\bar q_\b q_\a )_{V+A} \\
\oi_{7\g} &=& {e\over 4 \p^2} m_b \bar s_L \s^{\m\n} b_R F_{\m\n} \\
\oi_{8g} &=& {g_s\over 4 \p^2} m_b \bar s_L \s^{\m\n} T^a b_R G_{\m\n}^a \; .
\eea
In \eq{ham}, we did not write the operators $\oi_{1,2}^u$ obtained by
the replacements $c\to u$ in $\oi_{1,2}$, and the semileptonic
current--current operators which induce the transitions $b\to s \ell^+
\ell^-$. The effective Hamiltonian for $b\to d$ transitions can be
obtained via the substitution $s\to d$. Also, operators with different
helicity structures are not explicitely written and can be obtained
from the above operator basis via the replacement $L \leftrightarrow
R$~\cite{bghw}. We will comment on their impact on the numerical
analysis in section~\ref{sec4}.

The SM values of the Wilson coefficients $C_{1-10} (m_b)$ and
$C_{7\g,8g} (m_b)$ are given in table~\ref{wc110sm}; the opposite
chirality operators do not get standard model contributions if the
light masses are neglected.
\begin{table}[t]
\begin{center}
\begin{tabular}{|cc|cc|} \hline
$C_1$& -0.171   & $C_7$     & -0.00001 \cr 
$C_2$&  1.070   & $C_8$     &  0.0005 \cr
$C_3$& 0.0114   & $C_9$     & -0.01 \cr 
$C_4$& -0.0321  & $C_{10}$  &  0.0002 \cr
$C_5$&  0.00925 & $C_{7\g}$ & -0.313 \cr
$C_6$& -0.0383  & $C_{8g}$  & -0.188 \cr \hline
\end{tabular}
\caption{\it SM Wilson coefficients at the scale $m_b$.}
\label{wc110sm}
\end{center}
\end{table}

We will use the following approximate form~\footnote{This expression is
based on naive factorization and is therefore not exact;
but it correctly describes the way new physics enters.}
for the $B\to (J/\Psi,\phi) K$ amplitudes~\cite{fun}:
\bea
{\cal A}^{\psi} &=& - \sqrt{2} G_F \; f_\psi \; F_1^{B\to K}(m^2_\psi) \; m_\psi \; 
                    (\epsilon_\psi \cdot p_B) \; V_{tb}^{} V_{ts}^* \left[
                    C_\psi +{\a_s \over 2\p} {m_B^2\over m_\psi^2} C_{8g} r_8 
                    \tilde S_{\psi K} \right] \; , \\
{\cal A}^{\phi} &=& - \sqrt{2} G_F \; f_\phi \; F_1^{B\to K}(m^2_\phi) \; m_\phi \; 
                    (\epsilon_\phi \cdot p_B) \; V_{tb}^{} V_{ts}^* \left[
                    C_\phi + {8\over 9} P +
                    {\a_s \over 4\p} {m_b^2\over q^2} C_{8g} \tilde S_{\phi K} \right].
\label{ampl-phi}
\eea
Here, $f_{\psi(\phi)}$ is the decay constant of the $J/\Psi (\phi)$,
$F_1^{B\to K}(q^2)$ is the $B\to K$ penguin form factor, $\tilde
S_{(\phi,\psi) K}\simeq -0.76$ is the ratio of the chromo-magnetic to
penguin form factors for the $B\to (J/\Psi,\phi) K$ decay, $q^2 \simeq
m_b^2/2$, and $r_8\simeq 1/12$ is the ratio of colour octet and
singlet matrix elements. $P$ is an $O(\a_s)$ contribution to the
matrix elements of the QCD penguin operators, it is insensitive to new
short distance physics, and it is originated by loop diagrams with an
internal charm quark. Since the typical $q^2$ is above the charm
production threshold, $P$ carries a strong phase and its numerical
value is $-0.0132 - i\; 0.0145$~\cite{deshpande}. Finally, the
coefficients $C_{\psi,\phi}$ are
\bea
C_\psi &=& C_1 + C_3 + C_5 + {C_2 + C_4 + C_6\over 3} + 2 r_8 (C_2+C_4+C_6)\\
C_\phi &=& C_3 + C_4 + C_5 + {C_3 + C_4 + C_6\over 3} - {1\over 2} \left(
         C_7 + C_9 + C_{10} + {C_8 + C_9 + C_{10}\over 3} \right)
\eea
where all the $C_i$ are to be evaluated at the scale $m_b$.

New physics contribute to the above Wilson coefficients; the resulting
new phases will show up in the direct and mixing induced $CP$
asymmetries. Before writing an explicit model for such contributions,
we briefly discuss the renormalization group (RG) running of
$C_{\psi,\phi} (\m)$ from $\m= O(M_W)$ to $\m= O(m_b)$. The solution
of the NLO RG equations reads:
\bea 
C_\psi (m_b) &=& 0.3515 \left( 1+ 0.0018 \; R_3 - 0.0014 \; R_4 + 0.0020 \; R_5 +
                                0.0053 \; R_6 \right) \; ,\\
C_\phi (m_b) &=& -0.0243 \left( 1 -0.030 \; R_3 + 0.085 \; R_4
                                -0.028 \; R_5 + 0.044\; R_6 \right) 
\eea 
where $R_i \equiv C_i (M_W) / C_i^{SM} (M_W)$ and the other
coefficients are fixed to their SM value. It follows that the impact
of the QCD penguin matching conditions (the values of the Wilson
coefficients at the scale $M_W$) on $C_\psi (m_b)$ and $C_\phi (m_b)$
is respectively of order $1\%$ and $10\%$. Thus, as it is well known,
their effects on $C_\psi (m_b)$ is totally negligible: the
coefficients of the current-current operators totally dominate the
amplitude even if their contribution is colour suppressed. Thus, the
SM prediction of an extremely small direct $CP$ asymmetry in the
$J/\Psi$ mode is unaltered by new physics contributions. On the other
hand, new physics whose contributions are large enough to dominate the
phases of $C_{3-6}(M_W)$, may modify the overall phase of $C_\phi
(m_b)$ at the $10\%$ level. Taking into account that the $CP$
asymmetry depends on twice the phase of $C_\phi (m_b)$, we see that in
principle large ($O(0.2)$) deviations from the SM relation $a_{\psi
K}=a_{\phi K}$ are possible. Moreover, the presence of a term
proportional to $C_{8g}$ in \eq{ampl-phi} can have a strong impact if
there are large new complex contributions to the chromo-magnetic
dipole operator. In the standard model, this term is negligible.

Let us introduce the time-dependent $CP$ asymmetry in the decays
$B_d^0/\overline{B_d^0} \to \phi K$, given by
\beq
a_{\phi K}(t) \equiv \frac{\Gamma(B_d^0 (t) \to
 \phi K) - \Gamma(\overline{B_d^0}(t) \to \phi K)}
{\Gamma(B_d^0 (t) \to  \phi K) + \Gamma(\overline{B_d^0}(t) \to
 \phi K)} = {\cal A}_{\rm CP}^{\rm dir} \cos(\D M_{B_d} t) +
{\cal A}_{\rm CP}^{\rm mix} \sin (\D M_{B_d} t) \; ,
\label{aphiksdef}
\eeq
where $\D M_{B_d}$ is the is the $B_d$ -$\overline{B_d}$ mass
difference and $B_d^0(t)\; (\overline{B_d^0}(t))$ is the state at time
$t$ which started as a pure $B_d^0\; (\overline{B_d^0})$ at $t=0$.
${\cal A}_{\rm CP}^{\rm dir}$ and ${\cal A}_{\rm CP}^{\rm mix}$ are
the direct and mixing-induced CP-asymmetries, respectively. Their
explicit expressions are
\bea
{\cal A}_{\rm CP}^{\rm dir} &=& {1 -|\lambda_{\phi K}|^2 \over 1 + |\lambda_{\phi K}|^2} \\ 
{\cal A}_{\rm CP}^{\rm mix} &=& {2 \hbox{Im} \lambda_{\phi K} \over 1 + |\lambda_{\phi K}|^2} 
\eea
with 
\beq
\lambda_{\phi K_S} = e^{2 i (\b + \theta_d)} {\bar A \over A} \equiv 
                     e^{2 i (\b + \theta_d + \theta_A)} \left|\bar A \over A\right|.
\eeq
Here, $\b$ is the inner angle of the unitarity triangle of the
standard model, $\theta_d$ is a possible new physics contribution to
the phase of the $B_d - \bar B_d$ oscillations, and $A$ ($\bar A$) are
the amplitudes of the decay in question (and its $CP$ conjugate).

We turn next to the branching ratio and $CP$ asymmetry in the decay
$B\to X_s \g$ and to the $B_s$-$\bar B_s$ mass difference. These
observables are strongly affected by new physics in the FCNC $b\to s$
transition and therefore must be properly accounted for.

\subsection{$B\to X_s \g$}
The inclusive transition $B\to X_s \g$ plays a major role in limiting
possible new physics contributions to $B$ decays both through
constraints on the branching ratio and on direct $CP$ asymmetry. The
experimental information on the latter quantities
is~\cite{bsgexp}
\bea
\branch (B\to X_s \g) &=& (3.22 \pm 0.40) \times 10^{-4} \\
{\cal A}_{CP} (B\to X_s \g) &=& (- 3.5 \pm 7.7) \; \% \; .
\eea
In the numerical analysis we will use the NLO computation of $\branch
(B\to X_s\g)$ which, as a function of the Wilson coefficients
evaluated at the scale $M_W$ reads~\cite{kagan-neubert}:
\beq
\branch = \left[1.258 + 0.382 |R_7|^2  + 0.015 |R_8|^2 + 
          1.395 \; \hbox{Re} R_7 + 0.161 \; \hbox{Re} R_8 +
          0.083 \; \hbox{Re} (R_7 R_8^*) \right] 10^{-4}
\eeq
where $R_{7,8} = C_{7\g,8g}^{\rm tot}(M_W)/C_{7\g,8g}^{SM}(M_W)$.  The
parameter $\d$ that determines the cut on the photon energy spectrum
is set to 0.90 according to \rf{kagan-neubert}.   

Similarly, the $CP$ can be written as~\cite{kagan-neubert1}:
\beq
{\cal A}_{CP} = 1.06 \; \hbox{Im} {C_2 (m_b)\over C_{7\g} (m_b)} - 
                9.52 \; \hbox{Im} {C_{8g} (m_b)\over C_{7\g} (m_b)} +
                0.16 \; \hbox{Im} {C_2^{} (m_b) C_{8g}^* (m_b)\over |C_{7\g} (m_b)|^2} \; .
\eeq

\subsection{$\D M_{B_s}$}
The effective Hamiltonian for the $\D S=2$ transitions can be written
as
\beq
    {\cal H}_{eff}^{\D S=2}
=
    -\frac{G_F^2 M_W^2}{(2 \pi)^2} (V_{tb}^{} V_{ts}^*)^2
    \left[ C^{VLL} \oi^{VLL} +  \sum_{i=1}^2 \left(C^{SLL}_i \oi^{SLL}_i +
           C^{SLR}_i \oi^{SLR}_i \right) \right]+ {\it h. c.} \; .
\eeq
where
\bea
\oi^{VLL}   &=& (\bar s_L \g^\m b_L) (\bar s_L \g_\m b_L)   \; , \\
\oi^{SLL}_1 &=& (\bar s_L b_R) (\bar s_L b_R) \; , \\
\oi^{SLL}_2 &=& (\bar s^\alpha_L b^\b_R) (\bar s^\b_L b^\a_R) \; , \\
\oi^{SLR}_1 &=& (\bar s_L b_R) (\bar s_R b_L) \; , \\
\oi^{SLR}_2 &=& (\bar s^\alpha_L b^\b_R) (\bar s^\b_R b^\a_L) \; , \\
\eea
together with the operators $\oi^{VRR}$ and $\oi^{SRR}_{1,2}$,
obtained from the corresponding $LL$ ones via the substitution
$L\leftrightarrow R$.

In order to minimize the impact of hadronic uncertainties we will
consider, as usual, the ratio $X_{sd} \equiv \D M_{B_s} / \D
M_{B_d}$. Its explicit expression is~\cite{masiero}:
\beq 
X_{sd} = \xi^2 {M_{B_s}\over M_{B_d}} \left|
         V_{ts} \over V_{td} \right|^2 
         {{\cal C}^s \over {\cal C}^d }  \; .
\label{exs}
\eeq
where
\beq
{\cal C}^s =  C^{VLL} -{5 \kappa_s \over 8} C^{SLL}_1 + {\kappa_s \over 8} C^{SLL}_2
              + {6 \kappa_s +1\over 8} C^{SLR}_1 +{2 \kappa_s +3\over 8} C^{SLR}_2
\label{cs}
\eeq
with $\kappa_s = M_{B_s}^2/(m_b+m_s)^2$ and $\xi$ is a ratio of
hadronic matrix elements, numerically equal to $1.16 \pm
0.05$~\cite{smod}. The coefficient ${\cal C}^d$ is obtained by
replacing $s$ with $d$ in \eq{cs}. It is important to stress that the
contributions from new physics are generally different for the $B_s$
and $B_d$ systems. The ratio ${\cal C}^s / {\cal C}^d$ can therefore
be sizeably different from unity and has to be taken into account in
model independent analyses of the unitary triangle when using
\eq{exs}.

In the numerical analysis we will require $X_{sd}$ to lay in the
interval $(30\div 60)$ in order to satisfy the lower bound $\D M_{B_s}
> 14.9 \; \hbox{ps}^{-1}$ and have an observable $B_s$-$\bar B_s$ mass
difference at the same time.
\section{An example: SUSY gluino contributions}
\label{sec3}
In order to present a definite new physics model which contributes
sizeably to the $CP$ asymmetries in $B\to \phi K$, we turn to a
variant of the MSSM with complex off-diagonal squark mass terms. In
particular we will focus on the following entries:
$(m^2_{LL})^d_{23}$, $(m^2_{LR})^d_{23}$, $(m^2_{RR})^d_{23}$ and
$(m^2_{RL})^d_{23}$. The other parameters of the model are the common
mass of the squarks $\tilde m$, and the gluino mass $m_{\tilde
g}$. Following a common practice, we consider the normalized
insertions $\d^d_{23}$ given by the ratios of the various
$(m^2)^d_{23}$ to $\tilde m^2$.

It is well known, and in the next section we will give a detailed
quantitative analysis of the question, that $\branch (B\to X_s \g)$
puts severe bounds of order $O(10^{-2})$ on the $LR$ and $RL$
insertions while the impact on the $LL$ and $RR$ ones is rather
mild. This strong result follows from the $m_{\tilde g}/m_b$ chiral
enhancement of the $(\d^d_{23})_{LR,RL}$ contributions to the Wilson
coefficient $C_{7\g,8g}$. Note that this chiral factor is absent in
the in the WC's that govern the $B_s$-$\bar B_s$ mass difference and
the $B\to \phi K$ amplitude. For the latter observables, in fact, all
the mass insertions enter with similar weight: therefore, the only
insertions that can play a role are the $LL$ and $RR$ ones. Since the
analyses for both insertions give the same results, we will consider
explicitely only $(\d^d_{23})_{LL}$.

The expressions of the SUSY contributions to the various coefficients
are~\cite{ahrib}
\bea
C_3 &=& {\a_s^2  (\d^d_{23})_{LL}  \over 2\sqrt{2} G_F \tilde m^2 |V_{tb}^{} V_{ts}^*|}
        \left[ -{1\over 9} B_1 (x_{\tilde g \tilde q}) - {5\over 9} B_2 (x_{\tilde g \tilde q})
               -{1\over 3} P (x_{\tilde g \tilde q}) \right] \; , \\
C_4 &=& {\a_s^2  (\d^d_{23})_{LL}  \over 2\sqrt{2} G_F \tilde m^2 |V_{tb}^{} V_{ts}^*|}
        \left[ -{7\over 3} B_1 (x_{\tilde g \tilde q}) + {1\over 3} B_2 (x_{\tilde g \tilde q})
               + P (x_{\tilde g \tilde q}) \right] \; , \\
C_5 &=& {\a_s^2  (\d^d_{23})_{LL}  \over 2\sqrt{2} G_F \tilde m^2 |V_{tb}^{} V_{ts}^*|}
        \left[ {10\over 9} B_1 (x_{\tilde g \tilde q}) + {1\over 18} B_2 (x_{\tilde g \tilde q})
               -{1\over 3} P (x_{\tilde g \tilde q}) \right] \; , \\
C_6 &=& {\a_s^2  (\d^d_{23})_{LL}  \over 2\sqrt{2} G_F \tilde m^2 |V_{tb}^{} V_{ts}^*|}
        \left[ -{2\over 3} B_1 (x_{\tilde g \tilde q}) + {7\over 18} B_2 (x_{\tilde g \tilde q})
               + P (x_{\tilde g \tilde q}) \right] \; , \\
C_{7\g} &=& - {\pi \a_s (\d^d_{23})_{LL} \over \sqrt{2} G_F \tilde m^2 |V_{tb}^{} V_{ts}^*|} {16\over 9} 
            g_2 (x_{\tilde g \tilde q}) \; , \\
C_{8g}  &=& - {\pi \a_s (\d^d_{23})_{LL} \over \sqrt{2} G_F \tilde m^2 |V_{tb}^{} V_{ts}^*|} 
            \left( {1\over 3} g_2 (x_{\tilde g \tilde q}) + 3 g_1 (x_{\tilde g \tilde q}) \right)\; , \\
C^{VLL} &=& {\a_s (\d^d_{23})_{LL} \over \tilde m^2 |V_{tb}^{} V_{ts}^*|^2} \left(
            {1\over 9} x_{\tilde g \tilde q} f_6 (x_{\tilde g \tilde q}) + {23\over 72} 
             \tilde f_6 (x_{\tilde g \tilde q}) \right) \; , 
\eea
where $x_{\tilde g \tilde q} \equiv m^2_{\tilde g} / m^2_{\tilde q}$
(we assume a common squark mass).
The loop functions can be found in \rfs{masiero,ahrib}.  Note
that we have included the phase of $V_{tb}^{} V_{ts}^*$ (which is
basically a minus sign) in the definition of the mass insertion.

\section{Numerical analysis}
\label{sec4}
\begin{figure}[ht]
\begin{center}
\epsfig{file=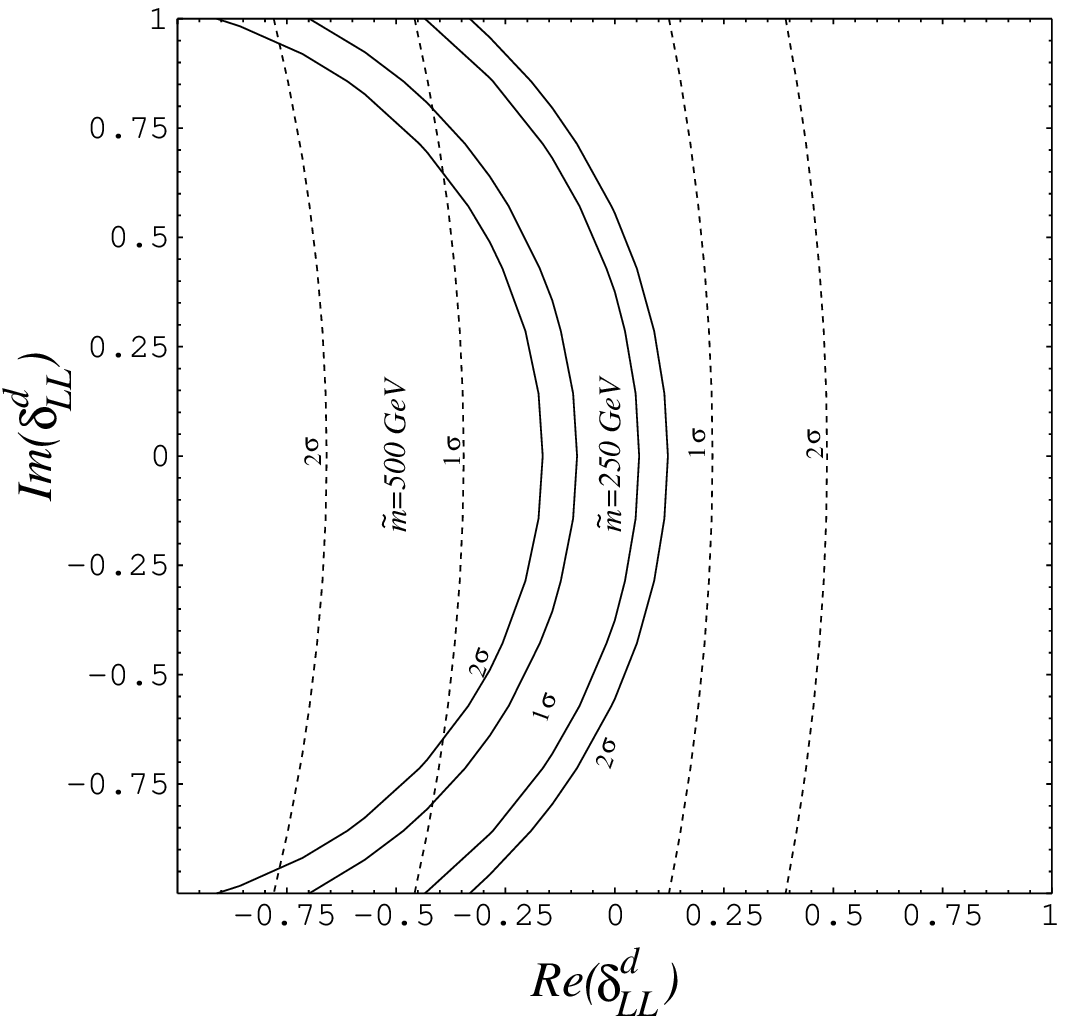,width=0.48\linewidth}
\epsfig{file=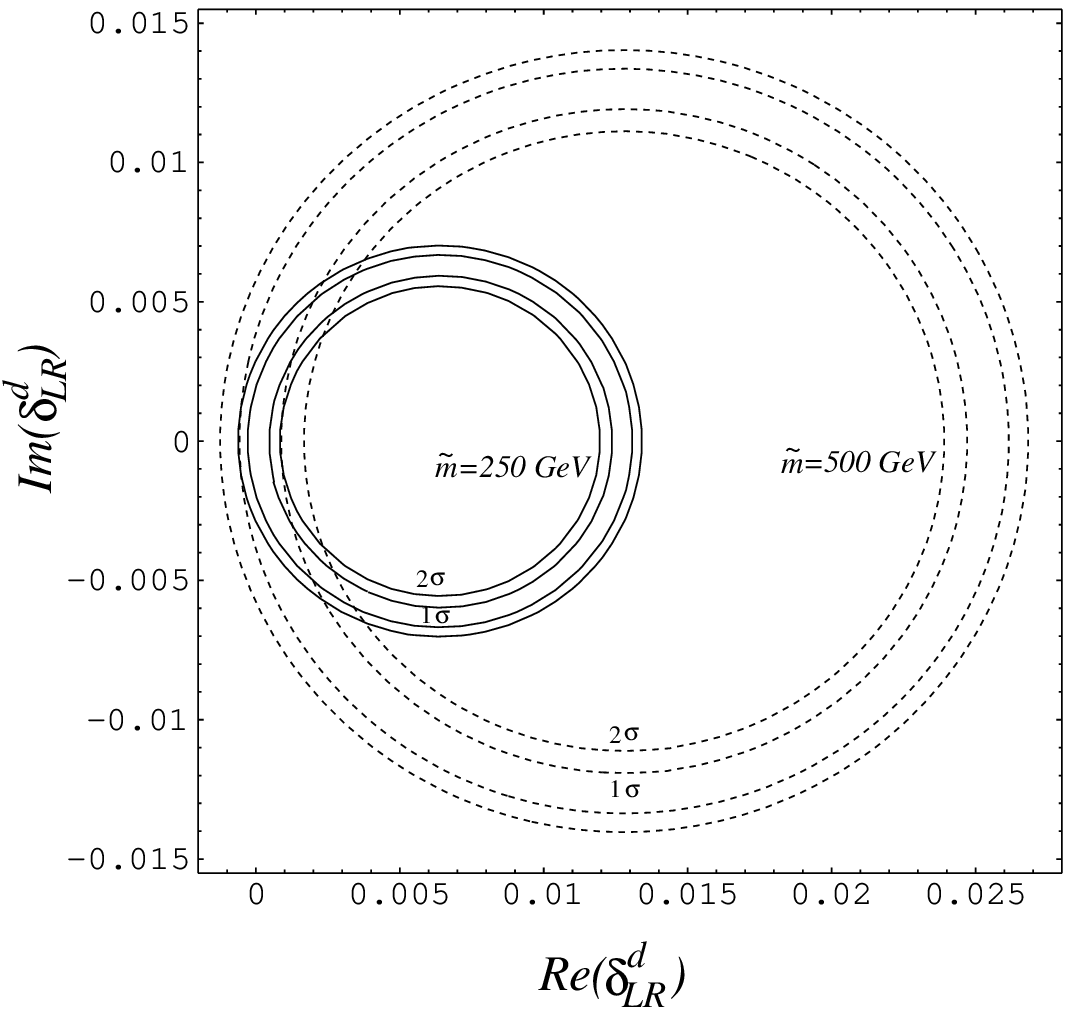,width=0.48\linewidth}
\caption{\it Constraints on the complex mass insertions 
$(\d^d_{23})_{LL}$ and $(\d^d_{23})_{LR}$ coming from $\branch (B\to
X_s\g)$. We consider $m_{\tilde g}^2 / \tilde m^2 =1$. The contours
scale as $\tilde m^2$. In the plot we explicitly show the cases
$\tilde m=250,500 \gev$.}
\label{bsg}
\end{center}
\end{figure}

We begin with the constraints imposed by the $B\to X_s \g$ branching
ratio and $CP$ asymmetry. We assume the absence of so-called
accidental cancellations between different large contributions;
therefore, we present the analysis assuming the presence of only one
insertion per time besides the SM. If all of the insertions are
substantial at the same time, the allowed ranges of some combination
of them can be sizeably enlarged (see for instance \rf{bhg} where
the case of multiple real mass insertions is considered).
In \fig{bsg} we take $x_{\tilde g \tilde q} = 1$, $\tilde m = 250, \;
500 \; \gev$ and require $|\d| < 1$; we also show the impact of using
the $\cl{68}$ and $\cl{95}$ constraints on the $B\to X_s \g$ branching
ratio. The corresponding bounds induced by the $CP$ asymmetry are much
weaker and do not impact, at the moment, the allowed areas: future
experimental improvements can substantially modify this picture. We
plot the resulting regions in the $[\hbox{Re} \d, \hbox{Im} \d]$
plane. Note that the contours scale with $\tilde m^2$.
As mentioned before, the bounds on the $LL$ insertion are not very
strong.  Even for very light squarks (i.e. $\tilde m = 250 \; \gev$)
it is possible to largely evade the constraint if the imaginary part
of the insertion is $O(1)$. This reflects the absence of interference
with the (real) SM contribution. On the other hand, the corresponding
bounds on the $LR$ insertion are, as expected, of order $O(10^{-2})$;
moreover, as it follows from the figure, if $|\d_{LR}|$ is larger than
$O(10^{-3})$, a substantial correlation between real and imaginary
parts is required.
For the $\d^d_{RR}$ and $\d^d_{RL}$ insertions the results are
similar.

\begin{figure}[ht]
\begin{center}
\epsfig{file=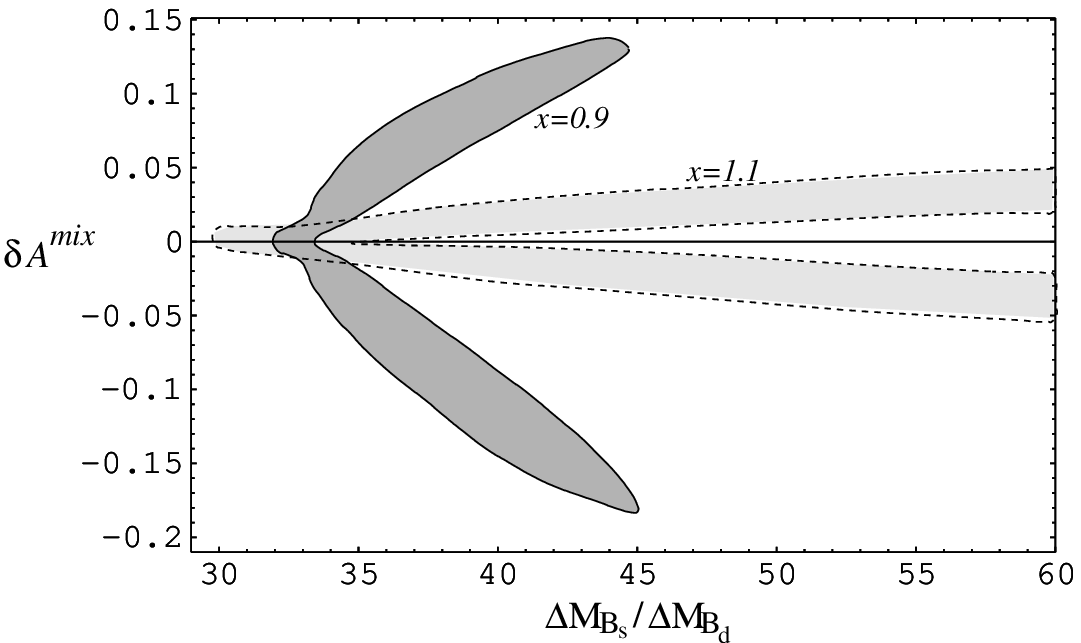,width=0.48\linewidth} 
\epsfig{file=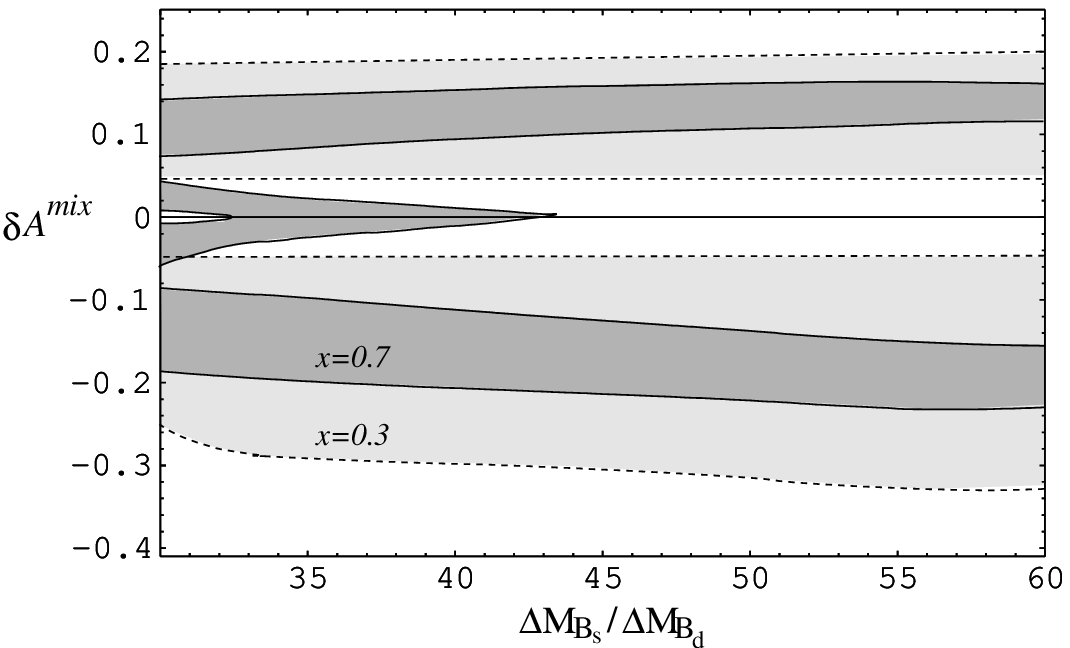,width=0.48\linewidth} 
\caption{\it Correlation between the $B_s$--$\bar B_s$ mass difference
and the possible deviations $\d A^{mix}$ of the mixing--induced $CP$
asymmetry from the SM expectation. The various regions correspond to
different values of $x_{\tilde g \tilde q}$.}
\label{amix}
\end{center}
\end{figure}

Next we analyze the possible deviations of the $CP$ asymmetries in
$B\to \phi K$ from the SM expectation and the relation to the
$B_s$-$\bar B_s$ mass difference which is also altered by the
insertions. In \fig{amix} we plot the correlation between the
$B_s$-$\bar B_s$ mass difference and $\d A^{mix}\equiv A^{mix} -
A^{mix,SM}$. We scan over the input SUSY parameters in the ranges
\bea 
&& \tilde m \in [250,1000] \gev \\ 
&& |(\d^d_{23})_{LL}| \in [0,1] \\ 
&& \arg (\d^d_{23})_{LL} \in [0,2 \p] \; , 
\eea
for different values of $x_{\tilde g \tilde q}$, the gluino-squark 
mass ratio. We require each point
to satisfy the $B\to X_s \g$ $\cl{95}$ constraint and to give a
$B_s$-$\bar B_s$ mass difference in the range $30 \leq X_{sd} \leq
60$. The various regions correspond to the limiting case: for a given
ratio $x_{\tilde g \tilde q}$, no point lays outside them. For
$x_{\tilde g \tilde q} \geq 1$ we find that the deviations from the
$SM$ expectation remain below 0.05. For smaller values of $x_{\tilde g
\tilde q}$ much larger and thus observable contributions are
possible. This strong dependence on the ratio between the gluino and
squark masses is due to the particular dependence of the loop
functions on $x_{\tilde g \tilde q}$. Moreover, the presence of
definite bands in the $[X_{sd}, \d A^{mix}]$ plane is due to an
interplay between the $B\to X_s \g$ constraint and the requirement of
fixed $\D M_{B_s}$.

As previously stated, $\d A^{mix}$ measures the deviation of the
mixing-induced $CP$ asymmetry from the standard model
prediction. Since all new physics in the $B_d$-$\bar B_d$
mixing amplitude will affect the $CP$ asymmetries
in the decays $B\to J/\Psi K$ and $B\to \phi K$ in the same way, 
$\d A^{mix}$ is equal to $\sin 2 (\b + \theta_d+
\theta_A) - \sin 2 (\b+\theta_d)$ where $\theta_d$ can be generated by
other SUSY couplings that we do not consider here and $\theta_A$ is
the phase of the $B\to \phi K$ decay amplitude. In other words,
$\d A^{mix}$ is the difference between the mixing-induced $CP$
asymmetries in the decays $B\to J/\Psi K$ and $B\to \phi K$ (see also
the discussion in \sec{sec2}).

\begin{figure}[ht]
\begin{center}
\epsfig{file=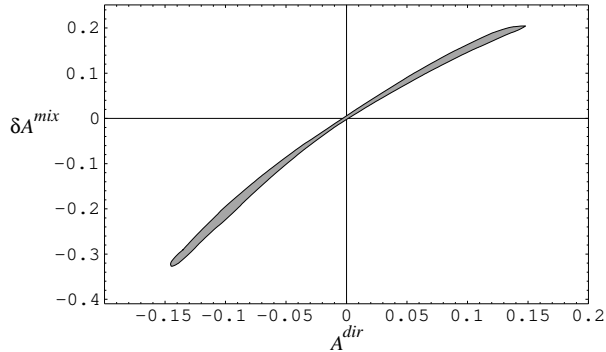,width=0.48\linewidth} 
\caption{\it Correlation between $\d A^{mix}$ and the direct $CP$
asymmetry $A^{dir}$.}
\label{adir}
\end{center}
\end{figure}

The new weak phases in the $B\to \phi K$ amplitude also
leads to a non vanishing direct $CP$ asymmetry, $A^{dir}$, which can
be measured for instance in decays of charged $B$-mesons. This
asymmetry depends crucially on the presence of a strong rescattering
phase, provided by the term $P$ of \eq{ampl-phi}. Nevertheless, we
find that the new physics contributions to the two asymmetries are
strongly correlated.  This can be understood as follows. Let us
parametrize the $B\to \phi K$ decay amplitude as $A_1 e^{i \phi_1}
e^{i \d_1}+ A_2 e^{i \phi_2} e^{i \d_2}$ where $\phi_i$ and $\d_i$ are
weak and strong phases respectively; note that this parametrization is
arbitrary but that all physical observables do not depend on its
choice. In the SUSY model that we consider, the weak phases $\phi_i$
are entirely due to the imaginary part of the mass insertions. Using
the above parametrization, we find that the ratio $A^{dir}/\d A^{mix}$
has an extremely tiny dependence on the phases $\phi_i$; therefore,
the correspondence between $A^{dir}$ and $\d A^{mix}$ is almost
one-to-one. In \fig{adir}, we explicitely show this correlation.
\section{Conclusions}
\label{concl}
We considered the effects of sizeable flavour changing entries in the
squark matrices on $b \to s$ transitions. In particular we allowed for
complex values of the relevant off-diagonal elements.

We first investigated the bounds that these entries must obey in order
to satisfy the $b \to s \gamma$ data. \fig{bsg} shows that the
inclusion of complex values for the mass insertion parameters strongly
enlarges their allowed regions. In fact, the absolute values of the
insertions can be much larger than in the existing
literature where only real couplings were considered. Moreover, there
are interesting correlations between real and imaginary parts which
may give important hints on the structure of the underlying theory.

We then considered the influence of the new terms on the
$CP$-violating asymmetry in the decays $B \to \phi K$ and on the
$B_s$-$\bar B_s$ mixing (the $B_d$-$\bar B_d$ mixing is not affected
by the terms we are interested in). In \fig{amix}, we plot the
deviation $\d A^{mix}$ of the mixing-induced $CP$ asymmetry in $B \to
\phi K$ from the SM expectation versus the ratio $X_{sd}$. We see that
$\d A^{mix}$ can reach the $20\%$ level in some corners of the
parameter space; on the other hand, deviations of order $O(10\%)$ are
easily possible with moderately light squark and gluino masses. Note
that such large contributions are possible only for configurations in
which $m_{\tilde g} \lesssim m_{\tilde q}$. In a similar fashion,
\fig{adir} shows that the direct $CP$ asymmetry $A^{dir}$ can receive
contributions of the same order of magnitude.  We stress that in this
framework effects on $B\to J/\Psi K$ decays are expected to be tiny.

The $B_s$-$\bar B_s$ $\D M_{B_s}$ is very sensitive to the mass insertions
we consider, while  $\D M_{B_d}$ remains unaffected.
This implies that the determination of $V_{td}$ from
ratio $\D M_{B_s}/\D M_{B_d}$ may be misleading. Moreover, as it
follows from \fig{amix} and \ref{adir}, an experimental determination
of $\D M_{B_s}$ in excess with respect to the SM prediction together
with sizeable $\d A^{mix}$ and $A^{dir}$ would be strong signatures in
favour of this kind of models. During the next year, the $B$-factories
BABAR and BELLE will gather enough luminosity to study the $CP$
asymmetries in $B \to \phi K$ decays and will test this class of SUSY
models soon.

\section*{Acknowledgments}
E.L. acknowledges financial support from the Alexander Von Humboldt
Foundation. D.W. is partially supported by Schweizerischer
Nationalfonds.

\end{document}